\documentclass[twocolumn,showpacs,prb,amsmath,amssymb]{revtex4}

\usepackage{graphicx}
\usepackage{dcolumn}
\usepackage{bm}

\usepackage{makeidx}
\usepackage{epsfig}
\usepackage{subfigure}
\usepackage{wrapfig}
\input comment.sty

\begin{document}

\font \tenrm=cmr10 at 8pt

\newcommand{\qed}{\nobreak \ifvmode \relax \else
      \ifdim\lastskip<1.5em \hskip-\lastskip
      \hskip1.5em plus0em minus0.5em \fi \nobreak
      \vrule height0.75em width0.5em depth0.25em\fi}

\title{Intrinsic optical bistability of thin films of linear
molecular aggregates: The one-exciton approximation}

\author{Joost A. Klugkist}
\author{Victor A. Malyshev}
\author{Jasper Knoester}
\affiliation{Center for Theoretical Physics and Zernike Institute
for Advanced Materials, University of Groningen, Nijenborgh 4, 9747
AG Groningen, The Netherlands }
\date{\today}

\begin{abstract}

We perform a theoretical study of the nonlinear optical response of
an ultrathin film consisting of oriented linear aggregates. A single
aggregate is described by a Frenkel exciton Hamiltonian with
uncorrelated on-site disorder. The exciton wavefunctions and
energies are found exactly by numerically diagonalizing the
Hamiltonian. The principal restriction we impose is that only the
optical transitions between the ground state and optically dominant
states of the one-exciton manifold are considered, whereas
transitions to other states, including those of higher exciton
manifolds, are neglected. The optical dynamics of the system is
treated within the framework of truncated optical Maxwell-Bloch
equations in which the electric polarization is calculated by using
a joint distribution of the transition frequency and the transition
dipole moment of the optically dominant states. This function
contains all the  statistical information about these two quantities
that govern the optical response, and is obtained numerically by
sampling many disorder realizations. We derive a steady-state
equation that establishes a relationship between the output and
input intensities of the electric field and show that within a
certain range of  the parameter space this equation exhibits a
three-valued solution for the output field. A time-domain analysis
is employed to investigate the stability of different branches of
the three-valued solutions and to get insight into switching times.
We discuss the possibility to experimentally verify the bistable
behavior.

\end{abstract}


\maketitle

\section{Introduction}

Optical circuits make use of light to process information. They
operate at the speed of light with almost no energy dissipation,
unlike electronic analogs. Optical
fibres~\cite{Mollenauer80,Hasegawa95} and photonic crystal
fibers~\cite{Russel03} have already found important applications in
optical communications and optoelectronic devices. Implementing
ultrafast optical sources and all-optical switches based on novel
(quantum-confined) materials, such as organic thin films and quantum
dots,~\cite{Wada04} as well as silicon-based
structures,~\cite{Almeida04} is now in progress. The realizability
of a single-photon optical switch based on warm rubidium vapor has
recently been demonstrated.~\cite{Dawes05}

A key element of any optical logic device is the optical switch,
which either passes or reflects the incoming light, depending on its
intensity. One possibility to design an optical switch is to utilize
the phenomenon of optical bistability. Since the theoretical
prediction of this effect by McCall~\cite{McCall74} and its
experimental demonstration by Gibbs, McCall, and
Venkatesan~\cite{Gibbs76} for a cavity filled with potassium atoms,
an extensive literature, both theoretical and experimental, has
developed on this topic (see
Refs.~\onlinecite{Abraham82,Lugiato84,Gibbs85} for historical
overviews and Ref.~\onlinecite{Rosanov96} for recent developments on
optical instability in wide aperture laser systems). A generic
optical bistable element exhibits two stable stationary transmission
states for the same input intensity, a property which in principle
opens the door to applications such as all-optical switches, optical
transistors, and optical memories.

Nonlinearity and feedback are the two necessary ingredients in order
to enable optical bistable response of an optical system. The former
can be provided, e.g., by a saturable medium, while a cavity
(mirrors) can serve to build up a feedback. This arrangement has
been used in the first demonstration of {\it controlling light with
light}.~\cite{Gibbs76} Sometimes, however, the nonlinearity itself
plays the role of the feedback. Here, bistability is an {\it
intrinsic} property of the material; no {\it external} feedback,
like a cavity, is needed. Thus, mirrorless (or cavityless) optical
bistability can be realized, which is even more advantageous from
the viewpoint of designing all-optical devices. During the past
decade, this type of bistability has been observed in a variety of
inorganic materials heavily doped with rare-earth
ions.~\cite{Hehlen94,Gamelin00,Noginov01,Goldner04} In
Refs.~\onlinecite{Hehlen94}, a population dependent dipole-dipole
interaction in ion pairs has been put forward as a nonlinearity and
feedback mechanism to explain the effect. This interpretation has
been debated in a number of papers.~\cite{Bodenschatz96,Guillot-Noel01,%
Malyshev98,Gamelin00,Noginov01,Goldner04,Ciccarello03}

Another class of materials, promising from the viewpoint of
all-optical manipulation of light, are molecular aggregates and
conjugated polymers. These systems commonly exhibit narrow
absorption bands and suppression of exciton-phonon coupling,
superradiance and giant optical nonlinearities, fast collective
optical response and efficient energy or charge transport (see for
an overview Refs.~\onlinecite{Spano94,Kobayashi96, Hadzii99,%
vanAmerongen00,Knoester02,Spano06,Scholes06}), which are ingredients
necessary to design optoelectronics or all-optical devices.
Molecular aggregates and conjugated polymers have already been used
to fabricate light emitting diodes~\cite{Greenham93} and organic
solid-state lasers.~\cite{Kranzelbinder00}

One particularly interesting effect, which has already received a
considerable amount of theoretical discussion, but still awaits
experimental realization, is the mirrorless optical bistability of a
single molecular aggregate~\cite{Malyshev96} or an assembly of
molecular aggregates.~\cite{Malyshev00,Jarque01,Glaeske01} The
bistable behavior of a {\it single} linear aggregate consists of a
sudden switching of the aggregate's excited state population from a
low level to a higher one upon a small change of the input intensity
around a critical point. The effect originates from a dynamic
resonance frequency shift, which depends on the number of excited
monomers in the aggregate. The origin of this shift lies in the
quasi-fermionic nature of Frenkel excitons in one
dimension.~\cite{Chesnut63,Agranovich68,Spano91} This nonlinearity
plays the role of {\it intrinsic} feedback, necessary for
bistability to occur. There exists, however, a restriction on the
aggregate length: an aggregate exhibits bistable behavior only if
its coherence length is larger than the emission wavelength, which
makes experimental realization problematic.

An {\it assembly} of molecular aggregates arranged in an ultrathin
film geometry (with the film thickness small compared to the
emission wavelength) may display intrinsic optical bistability
governed by another mechanism, where the density of molecules
becomes the driving parameter. The same mechanism holds for an
ultrathin film of homogeneously broadened two-level
systems.~\cite{Zakharov88} When the density in the film is high
enough, the on-resonance refractive index can get sufficiently large
to totally reflect an incoming field of low intensity. Then the
incoming field is almost completely compensated by a secondary field
of opposite phase, which is generated by the aggregate dipoles. The
dipole-induced field is bounded in magnitude, meaning that this
picture only holds if the incoming field intensity is smaller than a
certain value, determined by the density of aggregates. When this
value is exceeded, the aggregates become saturated, which suppresses
the dipole-induced field and abruptly changes the (nonlinear)
refractive index and transmittivity of the film. The field produced
by the aggregate dipoles plays the role of {\it intrinsic} feedback.
The output field depends nonlinearly on the input field of the film.

In Refs.~\onlinecite{Malyshev00}~-~\onlinecite{Glaeske01} a thin
film arrangement of oriented linear J-aggregates was considered,
where the localization segments of a single disordered aggregate
were modeled as independent homogeneous chains of fluctuating size.
Each segment was considered as a few-level system, with an
individual ground state and one or two excited states corresponding
to the dominant optical states of the segment. Within this
framework, both the ground state to one-exciton~\cite{Malyshev00}
and one-to-two~\cite{Glaeske01} exciton transitions were taken into
account, and bistable behavior was found in a certain region in the
parameter space.

The approach used in Refs.~\onlinecite{Malyshev00}
and~\onlinecite{Glaeske01} assumed full correlation of fluctuations
of the lowest exciton energy and the transition dipole moment,
taking both magnitudes as solely depending on the segment size. The
real picture, however, is quite
different.~\cite{Malyshev95,Malyshev01} In practise the optical
response of J-aggregates is strongly affected by disorder in the
molecular transition energies. The band-edge of the exciton energy
spectrum of such a disordered aggregate  is formed by states that
are localized on segments with small overlap. The lowest state of a
segment is optically dominant, whereas the other states have a much
smaller oscillator strength. The energy of the lowest state is not
correlated with the size of the segment; it is determined by
uncorrelated well-like fluctuations of the site
potential.~\cite{Lifshits68} Therefore, the optically dominant
states of non-overlapping segments can be arbitrarily close in
energy, having at the same time completely different transition
dipoles.~\cite{Bednarz04} In other words, the transition dipoles and
energies of the relevant states turn out to be uncorrelated rather
than correlated.

In this paper, we exploit the two-level model, implemented in
Refs.~\onlinecite{Malyshev00} and~\onlinecite{Jarque01}, to describe
the film's optical response. However, unlike
Refs.~\onlinecite{Malyshev00} and~\onlinecite{Jarque01}, we will
account properly for the statistical fluctuations of the transition
dipole moment and the transition energy, as they appear after
diagonalizing the Frenkel exciton Hamiltonian with uncorrelated
on-site disorder. We calculate the joint probability distribution of
these quantities  and use it to compute the electric polarization of
the film, which features in the Maxwell equation for the field. The
aggregate segment dynamics is described within the $2\times
2$-density matrix formalism. We derive a novel steady-state equation
for the output field intensity as a function of the input intensity
in terms of the joint probability distribution of the energy and the
transition dipole moment. On this basis, the bistability phase
diagram of the film is calculated. The critical parameter for
bistability to occur turns out to be different (larger) than that
found in Refs.~\onlinecite{Malyshev00}. By numerically solving the
truncated Maxwell-Bloch equations in the time domain, we study the
stability of the different branches of the three-valued solution for
the output field intensity. The calculation of an optical hysteresis
loop (an adiabatic up-and-down-scan of the field) demonstrates that
only two of them are stable. A new element in the paper is that we
also analyze switching time between both stable branches, and show
that it slows down dramatically close to the switching point.

The outline of this paper is as follows. In section~\ref{Sec: Model}
we present the model and mathematical formalism. Section~\ref{Sec:
Linear regime} deals with the linear regime of the transmission. The
steady state equation for the output intensity in the nonlinear
regime is derived in Sec.~\ref{Sec: Steady-state analysis}. In
Sec.~\ref{Sec: Time-domain analysis}, the stability of different
branches is considered, together with a study of the switching time.
In Sec.~\ref{Sec: Driving Parameters} we discuss the possibility to
achieve optical bistability using J-aggregates of polymethine dyes.
Section~\ref{Sec: Summary} summarizes the paper. Finally, in the
Appendix we address the effect of interference of the ground state
to one-exciton transitions, originating from the fact that excitons
are born from the same ground state, with all monomers being
unexcited.

\section{Model and formalism}
    \label{Sec: Model}

We aim to study the transmittivity of an assembly of linear
J-aggregates arranged in a thin film geometry (with the film
thickness $L$ small compared to the emission wavelength
$\lambda^{\prime}$ inside the film). All aggregates are aligned in
one direction, parallel to the film plane. Such an arrangement can
be achieved, e.g., by spin-coating.~\cite{Misawa93} The limit of $L
\ll \lambda^{\prime}$ allows one to neglect the inhomogeneity of the
field inside the film. The aggregates in the film are assumed to be
decoupled from each other. This finds its justification in the
strongly anisotropic nature of the system we have in mind. As we
will see later (Sec. VI), films of interest for bistability should
have a molecular density of the order of $10^{19}$ cm$^{-3}$. With a
typical separation of 1 nm between molecules within a single
aggregate, this implies that neighboring aggregates are separated by
10 nm. Thus, the dominant dipole-dipole interactions between
molecules of different chains are a factor of 1000 weaker than those
within chains. As a consequence, we expect that the former
interactions will merely result in small shifts of resonance
energies, away from the single-chain exciton energies considered
below.

On the other hand, the effect of interactions of the aggregate
molecules with the surrounding host molecules is important, because
as a consequence of the usually inhomogeneous nature of the host
media, they lead to disorder in the molecular transition energies
and in the molecular transfer integrals, both of which give rise to
localization of the exciton states on segments of the aggregates.
Finally, thermal fluctuations in the environment result in intraband
scattering of the excitons that causes two effects: equilibration of
the exciton population and homogeneous broadening of the exciton
levels. In this paper, we neglect the former effect. This finds its
justification in many experimental studies, which have shown that
the fluorescence Stokes shift of J-aggregates of cyanine dyes
usually is very small.~\cite{Fidder90,Minoshima94,Moll95,Kamalov96})

\subsection{A single aggregate}
    \label{Single_aggregate}

We model a single aggregate as a linear array of $N$ two-level
monomers with parallel transition dipoles. In this paper, we
restrict ourselves to optical transitions between the ground state
an the one-exciton manifold, described  by the Frenkel exciton
Hamiltonian
\begin{equation}
    H_0 = \sum_{n=1}^N \> \epsilon_n |n\rangle \langle n| +
    \sum_{n,m}^N\> J_{nm} \> |n\rangle \langle m| \ ,
\label{H}
\end{equation}
where $|n \rangle$ denotes the state in which the $n$th site is
excited and all the other sites are in the ground state and
$\epsilon_n$ is the excitation energy of site $n$. The $\epsilon_n$
are taken at random and uncorrelated from each other from a Gaussian
distribution with mean $\epsilon_0$ (the excitation energy of an
isolated monomer) and standard deviation $\sigma$. The transfer
interactions $J_{nm}$ are considered to be of dipolar origin and non
fluctuating: $J_{nm} = - J/|n-m|^{3}$ \, $(J_{nn} \equiv 0)$. Here
the parameter $J$ represents the nearest-neighbor transfer
interaction, which will be chosen positive (as is appropriate for
J-aggregates). The exciton energies $\varepsilon_\nu$ ($\nu =
1,\ldots , N$) and wavefunctions $|\nu\rangle = \sum_{n=1}^N
\varphi_{\nu n}|n\rangle$,  are obtained as eigenvalues and
eigenvectors of the $N \times N$ Hamilton matrix $H_{nm} = \langle
n| H |m \rangle$.

From the set of exciton states $|\nu^\prime\rangle$ we only take
into account the optically dominant states which, for $J > 0$,
reside in the neighborhood of the low-energy bare band edge at
$\varepsilon_0 = \epsilon_0 - 2.404 J$. These states are located at
different segments of the aggregate, which overlap weakly, and have
a wavefunction with no node. Therefore, they are called $s$-like
states. To find all such states, we use the selection rule proposed
in Ref.~\onlinecite{Malyshev01}. It reads \big|$\sum_n \varphi_{\nu
n} |\varphi_{\nu n}|\big| > C_0$, where we set $C_0 = 0.8$. This
rule selects states with a wavefunction consisting of mainly one
peak. From now on, the state index $\nu$ will count only such
$s$-like states. The number of these states is roughly equal to
$N/N^*$, where $N^*$ is their typical localization size. We assume
that the vibration-induced coherence length of excitons is much
larger than the disorder-induced localization length, a condition
that can be fulfilled at low temperature.~\cite{Heijs05} In this
limit, the exciton eigenstates $|\nu\rangle$ form a good basis.

The above picture implies that an aggregate is modeled as a
set of independent segments, each of which has its own ground
state $|0 \rangle$ and an $s$-like excited state $|\nu \rangle$.
The optical transition between these states is governed by the
segment dipole operator $\hat{d}_{\nu} = d_0 (|0\rangle \langle
\nu| + |\nu \rangle \langle 0|)$, where $d_0$ is the transition
dipole moment of a monomer. The corresponding transition dipole
moment of a segment is calculated as $d_{\nu} = d_0 \sum_{n}
\varphi_{\nu n} \equiv d_0 \mu_{\nu}$, where $\mu_{\nu} = \sum_{n}
\varphi_{\nu n}$ is the dimensionless transition dipole moment.

The optical dynamics of a segment is described in terms of the
$2\times 2$-density matrix ($\rho_{\nu\nu}, \rho_{\nu 0},
\rho^*_{\nu 0}, \rho_{00}$) which obeys the Bloch-like equations
(see the Appendix)
\begin{subequations}
    \label{Eq: Density matrix}
\begin{equation}
    \dot{\rho}_{\nu\nu} = -\gamma_{\nu} \rho_{\nu\nu}
    + i d_{\nu}\mathcal{E}\left(\rho_{0\nu}
    - \rho_{\nu 0}\right) \ ,
\end{equation}
\begin{equation}
    \dot{\rho}_{\nu 0} =
    -\left(i\varepsilon_{\nu} + \Gamma_{\nu} \right)\rho_{\nu 0}
    - i d_{\nu} \mathcal{E} (\rho_{\nu\nu} - \rho_{00})\ ,
\end{equation}
\begin{equation}
    \label{norma}
    \rho_{00} + \rho_{\nu\nu} = 1 \ .
\end{equation}
\end{subequations}
Here we set the Plank constant $\hbar = 1$ and introduced the
following notations: $\gamma_{\nu} = \gamma_0 |\mu_{\nu}|^2$ is the
radiative rate of the exciton state $\nu$ ($\gamma_0$ being the
monomer radiative rate), and $\Gamma_{\nu} = \frac{1}{2}
\gamma_{\nu} + \gamma_{\nu 0}$ is the dephasing rate of the state
$\nu$, which includes a pure dephasing term, $\gamma_{\nu 0}$.
Finally, $\cal{E}$ is the total electric field inside the film (see
below). Owing to the disorder, the transition energy
$\varepsilon_{\nu}$, the relaxation constant $\Gamma_{\nu}$, and the
transition dipole moment $\mu_{\nu}$ are stochastic variables, which
differ from segment to segment. Because of the fluctuations in
$\varepsilon_{\nu}$, $\Gamma_{\nu}$, and $d_{\nu}$, the density
matrix elements $\rho_{\nu\nu}$, $\rho_{\nu 0}$, and $\rho_{00}$
fluctuate as well.

\subsection{The Maxwell equation}
    \label{ME}

In this section, we specify the field $\mathcal{E}$ which enters
Eqs.~(\ref{Eq: Density matrix}). It consists of two contributions:
the incoming field $\mathcal{E}_i$ and a part produced by the
aggregate dipoles. The incoming field is considered to be a plane
wave $\mathcal{E}_i = E_i(x,t) \cos(k_i x - \omega_i t)$ with a
frequency $\omega_i$ and an amplitude $E_i(x,t)$, normally incident
and polarized along the aggregate transition dipoles. Under these
conditions, all the vectorial variables (transition dipole moments,
incoming and outgoing fields, and the field inside the film) can be
considered as scalars. The amplitude $E_i(x,t)$ is assumed to vary
slowly on the scale of the optical period $2\pi/\omega_i$ and
wavelength $\lambda_i = 2\pi/k_i$.

We assume without loss of generality that the film is located  in
the ZY plane ($x = 0$). Then the total field at $x = 0$ (inside the
film) is given by~\cite{Benedict88,Benedict96}
\begin{equation}
    \label{Eq:Maxwell for a thin film}
    \mathcal{E} = \mathcal{E}_i - \frac{2\pi L}{c}\dot{\mathcal{P}}
    \ ,
\end{equation}
where $\mathcal{P}$ is the electric polarization of the film
(electric dipole moment per unit volume), the dot denotes the time
derivative, and $c$ stands for speed of light. The second term in
the right hand side of Eq.~(\ref{Eq:Maxwell for a thin film})
represents the field produced by the dipoles in the film, emitted
perpendicular to the film in both directions. The part propagating
to the left is the reflected (plane wave) field, given at $x = 0$ by
$\mathcal{E}_{r} = -(2\pi L/c)\dot{\mathcal{P}}$, while the part
propagating to the right is the emitted (also plane wave) field,
which forms, together with the incident field $\mathcal{E}_i$, the
transmitted signal, determined at $x = 0$ by Eq.~(\ref{Eq:Maxwell
for a thin film}).

The electric polarization $\mathcal{P}$ is calculated as follows.
First, we introduce the expectation value of the dipole operator of
an aggregate, $d = d_0 \sum_{\nu \in s}\> \mu_{\nu} (\rho_{\nu 0} +
\rho_{0 \nu})$, where the summation is performed only over the
$s$-like states of the aggregate. Furthermore, this value is
averaged over a physical volume $V$, containing $M$ aggregates,
which, in fact, is equivalent to obtaining the average $\langle d
\rangle$ over disorder realizations. After that, ${\cal P}$ is
obtained by multiplying $\langle d \rangle$, with the number density
$M/V$ of the aggregates. The final formula for the electric
polarization reads:
\begin{equation}
    \label{P}
    \mathcal{P}
    = d_0 n_0 \frac{N_s}{N}\int d\varepsilon d\mu \,
    \mathcal{G}_s(\varepsilon,\mu)\, \mu \,
    [\, \rho_{10}(\varepsilon,\mu,t) + \mathrm{c.c.}] \ .
\end{equation}
Here, $n_0 = NM/V$ is the number density of monomers, $N_s =
\Big\langle \sum_{\nu \in s} \> 1 \Big\rangle$ a normalization
constant (having the meaning of the average number of $s$-like
states in an aggregate), and $\rho_{10}(\varepsilon,\mu,t)$ is the
off-diagonal density matrix element, where the indices 0 and 1 label
the ground and the excited $s$-state of the segment, respectively.
In our present formulation this element, as well as $\rho_{00}$ and
$\rho_{11}$, are ordinary (not stochastic) functions of
$\varepsilon$ and $\mu$; which formally follow from solving
Eqs.(\ref{Eq: Density matrix}). All stochastic aspects of the
segment's properties are taken into account through the function
$\mathcal{G}_s(\varepsilon,\mu)$, which represents the joint
probability distribution of the transition energy $\varepsilon$ and
the dimensionless transition dipole moment $\mu$ of the segment. The
latter is defined as
\begin{equation}
\label{G}
    \mathcal{G}_s(\varepsilon,\mu) = \frac{1}{N_s} \left\langle
    \sum_{\nu \in s} \delta\Big(\varepsilon - \varepsilon_{\nu}\Big)
    \delta \Big(\mu - \mu_{\nu} \Big)\right\rangle.
\end{equation}
It is worth to notice that at a given disorder strength $\sigma$,
$N_s$ scales linearly with the aggregate size $N$. Hence, the ratio
$N_s/N$ in Eq.~(\ref{P}) is $N$-independent. From our simulations we
found that $N_s/N = 0.074 (\sigma/J)^{0.8}$, which nicely agrees
with the disorder scaling of the typical localization size
$N^*$.~\cite{Malyshev01}


\begin{figure}
\begin{center}
\includegraphics[width = .48\textwidth,scale=1]{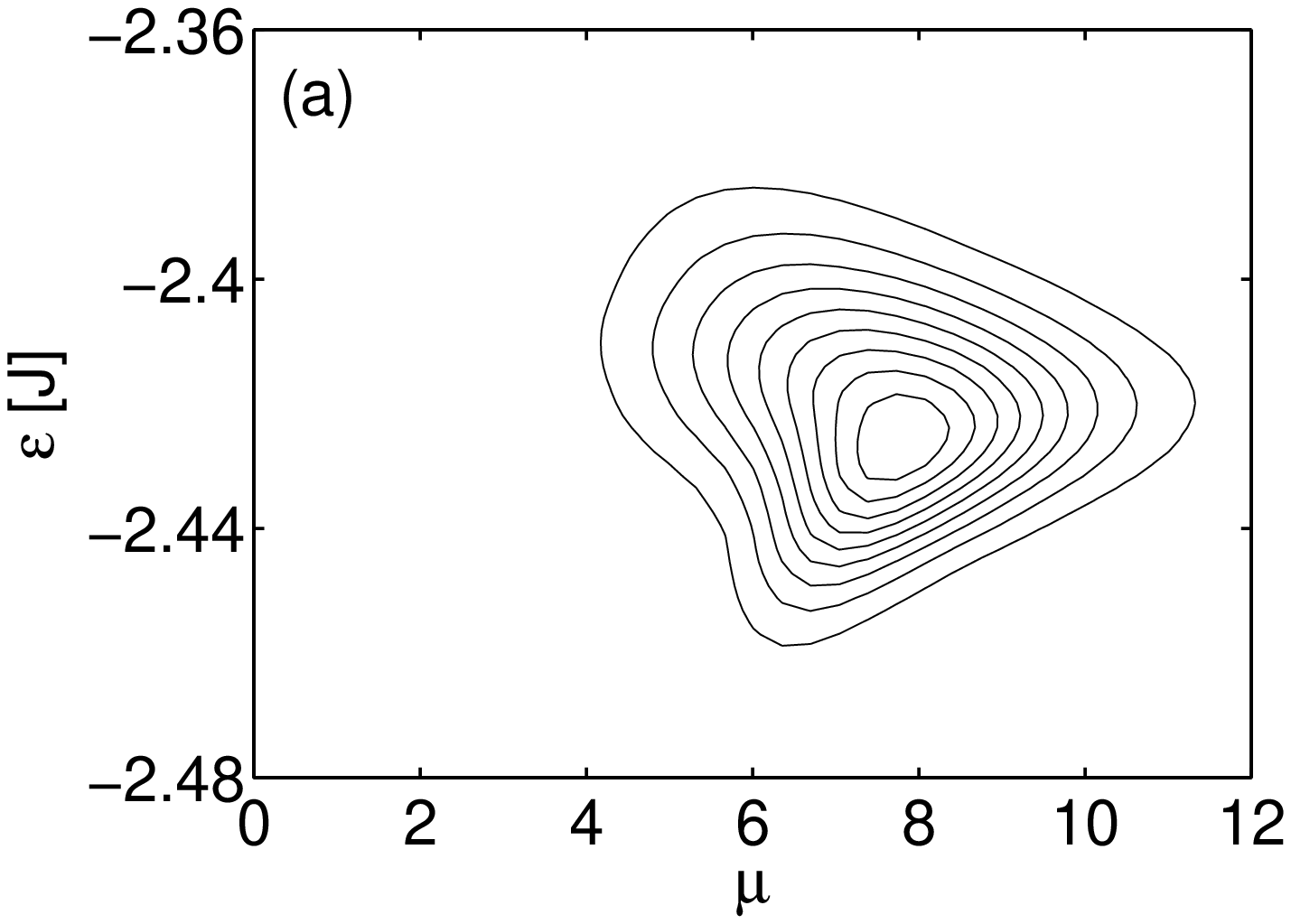}
\end{center}
\begin{center}
\includegraphics[width = .48\textwidth,scale=1]{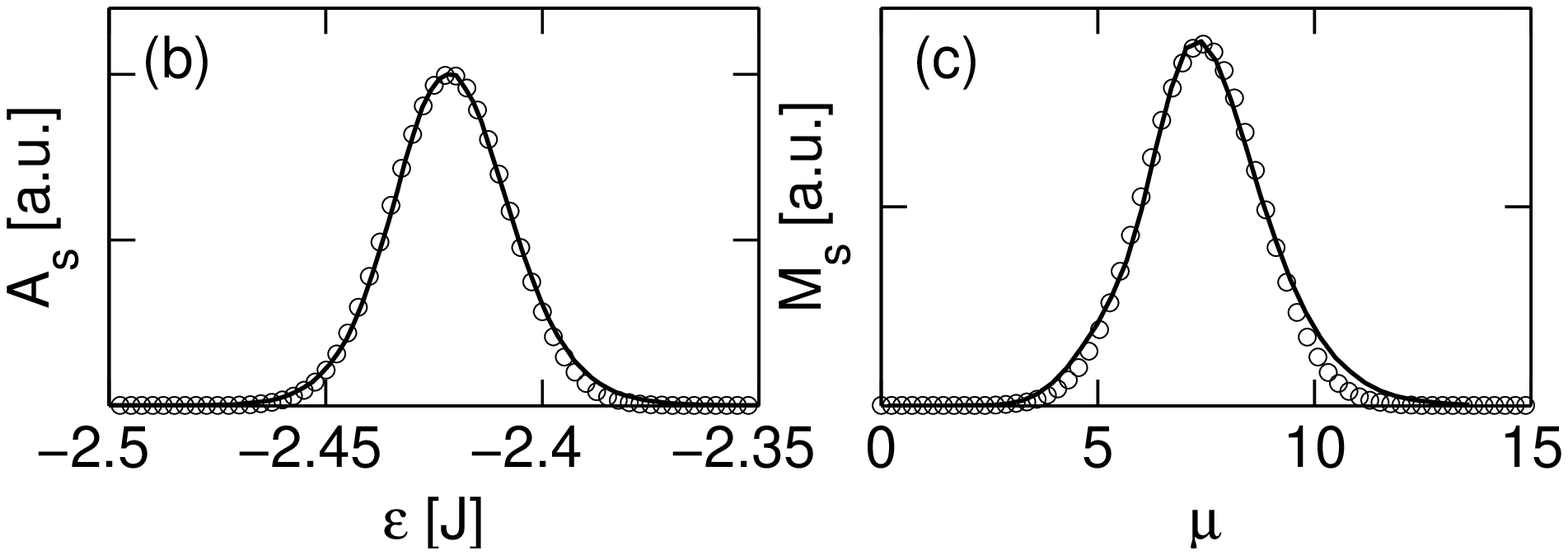}
\end{center}
\caption{(a) The joint probability distribution
    $\mathcal{G}_s(\varepsilon,\mu)$ of the transition energy
    $\varepsilon$ and dimensionless transition dipole moment $\mu$
    for $s$-like states on localization segements, obtained for
    a disorder strength $\sigma = 0.1 J$ according to Eq.~(\ref{G}).
    We used chains of length $N=500$ with the monomer transition
    energy $\epsilon_0 = 0$. The sampling was performed over
    300 000 disorder realizations. Contour lines correspond to
    10\% of the peak value of the distribution.
    (b) - The absorption spectrum $\mathcal{A}_s(\varepsilon)
    = \int d\mu \> \mu^2 \mathcal{G}_s(\varepsilon,\mu)$.
    (c) - The distribution $\mathcal{M}_s(\mu) = \int d\varepsilon \,
    \mathcal{G}_s(\varepsilon,\mu)$ of the transition dipole
    moment $\mu$. The solid lines represent the results of
    calculations, whereas the open circles are fits by a Gaussian.}
\label{fig: Example G}
\end{figure}

After the ${\cal G}_s$-distribution is obtained by straightforward
sampling of a sufficient number of disorder realizations, one can
easily calculate the two important quantities: ${\cal
A}_s(\varepsilon) = N_s^{-1} \big\langle \sum_{\nu \in s}
\mu^2_{\nu} \delta\big(\varepsilon - \varepsilon_{\nu}\big)
\big\rangle = \int d\mu \> \mu^2 \, \mathcal{G}_s(\varepsilon,\mu)$,
which represents the absorption spectrum, not accounting for
homogeneous broadening (i.e., close to the zero-temperature
spectrum), and $\mathcal{M}_s(\mu) = N_s^{-1} \big\langle \sum_{\nu
\in s} \delta \big(\mu - \mu_{\nu} \big)\big\rangle = \int
d\varepsilon \, \mathcal{G}_s(\varepsilon,\mu)$, which represents
the probability density of the transition dipole moment. As we are
mostly interested in the limit of dominating ingomogeneous
broadening, we will refer from now on to ${\cal A}_s(\varepsilon)$
as to the absorption spectrum, assuming that its half width at half
maximum (HWHM) $\sigma^*$ is larger than the homogenous HWHM
(resulting from $\Gamma_{\nu}$).

An example of the distributions $\mathcal{G}_s(\varepsilon,\mu),
{\cal A}_s(\varepsilon)$, and $\mathcal{M}_s(\mu)$ computed for an
ensemble of chains with $N = 500$ and a disorder strength $\sigma =
0.1 J$, is depicted in Fig.~\ref{fig: Example G} [panels~(a),(b),
and (c), respectively].
%
%
Because $\mathcal{G}_s(\varepsilon,\mu) =
\mathcal{G}_s(\varepsilon,-\mu)$, only $\mu > 0$ is considered in
the plots.

Note that in our model, the absorption spectrum
$\mathcal{A}_s(\varepsilon)$ is almost symmetric with  respect to
the peak position, except the tails, which show a small asymmetry.
It can be fitted well by a Gaussian, unlike the case when all the
exciton states are taken into account. The latter gives rise to a
Lorentzian high-energy tail of $\mathcal{A}_s(\varepsilon)$,
reproducing the asymmetric lineshape commonly seen in experiments.
The shape of the $\mathcal{M}_s$-distribution can also be fitted by
a Gaussian, but with a lesser accuracy than the absorption spectrum.
The distribution $\mathcal{G}_s(\varepsilon,\mu)$ exhibits
interesting scaling properties with regard to the disorder strength
$\sigma$. A detailed study will be presented elsewhere.

\subsection{Truncated Maxwell-Bloch equations}
    \label{MBE}

To proceed we seek the solution of Eqs.~(\ref{Eq: Density matrix})
in the standard manner: we set $\rho_{10} = -(i/2) R
\exp{(-i\omega_i t)}$ and $\mathcal{E} = (1/2)E \exp{(-i \omega_i
t)} + \mathrm{c.c.}$, where the complex amplitudes $R$ and $E$ vary
slowly on the time scale $2\pi/\omega_i$, and we use the rotating
wave approximation. It is straightforward to arrive at a set of
truncated equations for the populations $\rho_{11}$ of the
one-exciton states, and the amplitudes $R$ of the off-diagonal
density matrix elements, and the field $\Omega = d_0 E$ (in
frequency units):
\begin{subequations}
    \label{Eq: Density matrix truncated}
\begin{equation}
    \dot{\rho}_{11} = -\gamma \rho_{11}
    - \frac{1}{4} \mu \left(\Omega R^* + \Omega^*R \right) \ ,
\end{equation}
\begin{equation}
    \dot{R} = -\left[i(\Delta - \Delta_0) + \Gamma \right]R
    + \mu \Omega (\rho_{11} - \rho_{00})\ ,
\end{equation}
\begin{equation}
    \Omega =\Omega_i  + \Gamma_R \> \frac{N_s}{N}\, \int d\Delta
    d\mu \, \mathcal{G}(\Delta,\mu)\, \mu R \ ,
\end{equation}
\end{subequations}
where $\Delta - \Delta_0 = \varepsilon - \omega_i$ is the frequency
detuning between the exciton transition and the incoming field,
which is decomposed into two parts: $\Delta = \varepsilon -
\varepsilon_0$ and $\Delta_0 = \omega_i - \varepsilon_0$ indicating,
respectively, the frequency detuning of the exciton state and the
incoming field from the exciton band-edge frequency $\varepsilon_0 =
\epsilon_0 - 2.404 J$.

The constant $\Gamma_R = 2\pi n_0 {d_0}^2 k L$ is an important
parameter of the model.~\cite{Malyshev00,Jarque01,Glaeske01} The
physical meaning of $\Gamma_R$ can be explored by rewriting it in
the form $\Gamma_R = \frac{3}{2\pi}\gamma_0 n_0 L (\lambda/2)^2$,
where $\gamma_0 = 4 d_0^2 \omega^3/(3c^3)$ is the monomer
spontaneous emission rate and $n_0 L$ is the surface density of
monomers. The quantity $n_0 L(\lambda/2)^2$ can be interpreted as
the number of monomers in a $(\lambda/2)^2$-square that oscillate in
phase. In other words, $\Gamma_R$ can be considered as the radiative
rate of a single monomer, $\gamma_0$, enhanced by the number of
monomers within a $(\lambda/2)^2$-square.~\cite{Lee74} $\Gamma_R$
governs the Dicke superradiance of a thin
film,~\cite{Benedict88,Benedict96} as well as the collective
radiative damping in the linear regime (see the next section).
Therefore it is usually referred to as the superradiant constant.

The set of equations~(\ref{Eq: Density matrix truncated}) together
with the normalization condition~(\ref{norma}) and the
definition~(\ref{G}) form the basis of our analysis. In the
remainder of this paper, we will be particularly interested in the
dependence of the transmitted field intensity $|\Omega|^2$ on the
input field intensity $|\Omega_i|^2$ following from these equations.

\section{Linear regime}
\label{Sec: Linear regime}

In order to get insight into the effect and interplay of the
parameters that govern the bistability, we first consider the linear
regime of the system. We assume that a weak input field $\Omega_i$ =
const is switched on at $t=0$, weakness implying that the depletion
of the ground state population can be neglected.   Thus, we set
$\rho_{00}(t) = 1$ and $\rho_{11}(t) = 0$, which linearizes
Eqs.~(\ref{Eq: Density matrix truncated}),
\begin{subequations}
    \label{Eq: Density matrix_truncated linear}
\begin{equation}
    \dot{R} = -\left[i(\Delta - \Delta_0) + \Gamma \right]R
    - \mu \Omega \ ,
\end{equation}
\begin{equation}
    \Omega =\Omega_i  + \Gamma_R \> \frac{N_S}{N}\int d\Delta  d\mu \,
    \mathcal{G}_s(\Delta,\mu)\, \mu R.
\end{equation}
\end{subequations}
These equations can be solved easily  in the Laplace domain. The
solution for the Laplace transform of the transmitted field
$\tilde{\Omega}$ reads
\begin{eqnarray}
    \label{Eq: Linear solution 1}
    \tilde{\Omega} = \Big[1 + \Gamma_R \> \frac{N_s}{N} \int d\Delta
    d\mu \, \mathcal{G}_s(\Delta,\mu)\, \mu^2
\nonumber\\
\nonumber \\
    \times \frac{1}{p + \left[i(\Delta
    - \Delta_0) + \Gamma \right]}\Big]^{-1}\> \tilde{\Omega}_i
    \ ,
\end{eqnarray}
where $p$ denotes the Laplace parameter. To evaluate this
expression, we neglect the $\mu$-dependence of $\Gamma$. Then the
integral over $\mu$ gives, by definition, the  absorption spectrum
$\mathcal{A}_s(\Delta)$. The latter is normalized now to $F_s/N_s$,
where $F_s = \big \langle \sum_{\mu \in s}\> \mu_{\nu}^2 \big
\rangle$ is the average total oscillator strength of the $s$-like
states per aggregate. To perform the integration over $\Delta$
explicitly, we replace $\mathcal{A}_s(\Delta)$ by a Lorentzian
centered at $\Delta^*$ and with  a width $\sigma^*$:
\begin{equation}
    \label{Eq: A Lorentzian}
    \mathcal{A}_s(\Delta) = \frac{F_s}{N_s} \> \frac{\sigma^*}{\pi}
    \> \frac{1}{\left[(\Delta - \Delta^*)^2 + {\sigma^*}^2 \right]}
\end{equation}
(in all our numerical results, we do not invoke this approximation
and keep the exact form of the ${\cal G}_s$-distribution, i.e., of
the absorption spectrum). With this substitution, the result of the
integration over $\Delta$ reads:
\begin{eqnarray}
    \label{Eq: Linear solution 2}
    \tilde{\Omega} = \tilde{\Omega}_i -  \frac{\tilde{\Gamma}_R}
    {p + i(\Delta^* - \Delta_0) + \Gamma + \sigma^*
    + \tilde{\Gamma}_R } \> \tilde{\Omega}_i \ ,
\end{eqnarray}
where we introduced the renormalized superradiant constant
$\tilde{\Gamma}_R = (F_s/N)\Gamma_R$. As the total oscillator
strength of $s$-like states $F_s < N$, the ratio $F_s/N < 1$. We
also note that $\Gamma + \sigma^*$ denotes the total (homogeneous
plus inhomogeneous) dephasing rate.

Finally, by assuming $\Omega_i = \text{const}$, which corresponds
to $\tilde{\Omega}_i = \Omega_i/s$ in the Laplace domain, we
obtain the following time-domain behavior of the transmitted field
\begin{eqnarray}
    \label{Eq: Linear solution 3}
    \Omega & = & \frac{i(\Delta^* - \Delta_0) + \Gamma + \sigma^*}
    {i(\Delta^* - \Delta_0) + \Gamma + \sigma^* + \tilde{\Gamma}_R} \>
    \Omega_i
\nonumber\\
\nonumber\\
    & +  & \frac{\tilde{\Gamma}_R}{i(\Delta^* - \Delta_0) + \Gamma
    + \sigma^* + \tilde{\Gamma}_R } \> \Omega_i
\nonumber\\
\nonumber\\
    & \times & \exp\left[ -i(\Delta^* - \Delta_0)t - (\Gamma +
    \sigma^* + \tilde{\Gamma}_R) t \right] \ .
\end{eqnarray}

As is seen from this equation, the field in the film, $\Omega$,
reaches its steady-state value (given by the first term in the
right-hand side) after a time $1/(\Gamma + \sigma^* +
\tilde{\Gamma}_R)$. If the dephasing dominates the relaxation of the
dipoles, i.e., if $\Gamma + \sigma^* \gg \tilde{\Gamma}_R$, the
steady state limit of the opposing dipole field, given by $-
\Omega_i\tilde{\Gamma}_R/[i(\Delta^* - \Delta_0) + \Gamma + \sigma^*
+ \Gamma_R]$, is small in magnitude compared to the incoming field
$\Omega_i$. As a consequence, the field inside the film $\Omega
\approx \Omega_i$. In this limit, one finds a \textit{high} film
transmittivity.

When $\tilde{\Gamma}_R \gg \Gamma + \sigma^*$  the superradiant
damping drives the relaxation. Now the film dipoles, having
sufficient time to respond collectively, can produce an opposing
field $- \Omega_i\tilde{\Gamma}_R/|i(\Delta^* - \Delta_0) + \Gamma +
\sigma^* + \tilde{\Gamma}_R|$ of magnitude $\approx \Omega_i$. This
field almost totally compensates the incoming field, and results in
a low magnitude of the field inside the film, $|\Omega| \sim
\Omega_i|i(\Delta^* - \Delta_0) +\Gamma + \sigma^*|/\tilde{\Gamma}_R
\ll \Omega_i$, and, consequently, in a \textit{low} film
transmittivity. Switching to a high transmission state now requires
a field intensity $\Omega_i$ that saturates the system. In this case
we can see optical bistable switching (see the next section).

From the above, it is clear that the interplay of superradiance and
dephasing determines the linear transmittivity of the film. Hence,
the ratio $\tilde{\Gamma}_R/(\Gamma + \sigma^*)$ is an important
parameter of the model. In the theory of bistability it is often
referred to as the \textit{cooperative
number}.~\cite{Lugiato84,Gibbs85}

\section{Steady-state analysis}
    \label{Sec: Steady-state analysis}

\subsection{Bistability equation}
    \label{Sec: Bistability equation}

In this section, we consider the steady-state regime, when we set
$\Omega_i = \text{const}$ and $\dot{R} = \dot{\rho}_{11} = 0$. It
is a matter of simple algebra to derive the following equation for
the output intensity $|\Omega|^2$:
\begin{eqnarray}
    \label{Eq: Steady state output field}
    \Omega_i^2 & = & |\Omega|^2 \Big| 1 + \Gamma_R \> \frac{N_s}{N}
    \int d\Delta d\mu \, \mu^2 \mathcal{G}_s(\Delta,\mu)
\nonumber\\
\nonumber\\
    & \times & \frac{\Gamma - i(\Delta - \Delta_0)}
    {(\Delta - \Delta_0)^2 + \Gamma^2 +
    |\Omega|^2 \Gamma/\gamma_0} \Big|^2 ,
\end{eqnarray}

Formally, Eq.~(\ref{Eq: Steady state output field}) differs from the
one found previously~\cite{Malyshev00} by the small factor $N_s/N$.
This smallness, however, is compensated by the $N_s$-scaling of the
integral in~(\ref{Eq: Steady state output field}): the latter is
proportional to $F_s/N_s \gg 1$ (see the preceding section). Thus,
the actual numerical factor in Eq.~(\ref{Eq: Steady state output
field}) is on the order of $F_s/N$. Numerically, we found that
$F_s/N$ depends only weakly on the disorder strength $\sigma$, lying
within an interval $0.75 \le F_s \le 0.83$ when the disorder
strength $\sigma$ ranges from 0 to $0.5J$. This means that the
linear optical response in a system with static disorder is
dominated by the $s$-like states, independent of the disorder. We
stress that, unlike previous works,~\cite{Malyshev00} Eq.~(\ref{Eq:
Steady state output field}) properly  accounts for the joint
statistics of the transition energy and the transition dipole moment
via the $\mathcal{G}_s$-distribution.

\subsection{Phase diagram}
    \label{Sec: Phase diagram}

Numerical analysis shows that Eq.~(\ref{Eq: Steady state output
field}) can have three real roots in a certain region of the
parameter space $(\Gamma_R,\sigma^*)$. In other words, our model can
exhibit bistable behavior. In all simulations, we used linear chains
of $N = 500$ sites and $\gamma_0 = 2\times 10^{-5}J$ (appropriate
for monomers of polimethine dyes). The dephasing constant
$\gamma_{\nu 0}$ was considered not fluctuating~\cite{Heijs05} and
was set to $\gamma_{\nu 0} = 500 \gamma_0$.

\begin{figure}
\begin{center}
\includegraphics[width = 0.45\textwidth,scale=1.1]{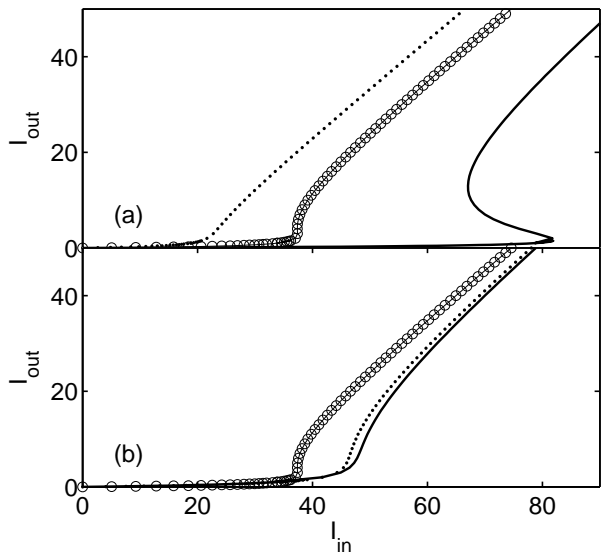}
\end{center}
\caption{Examples of the input-output characteristics, demonstrating
the occurrence of three-valued solutions to Eq.~(\ref{Eq: Steady
state output field}). In simulations, chains of $N = 500$ sites and
a disorder strength $\sigma = 0.1 J$ were used, corresponding to a
HWHM $\sigma^* = 0.0156 J$.
    (a) - The results obtained for different superradiant constants
    $\Gamma_R$ at the optimal detuning $\Delta_0^{\mathrm{opt}} =
    -2.42 J$, which corresponds to an incoming field which is resonant
    with the absorption maximum. The open circles, dotted, and solid
    curves represent, respectively, the data calculated for $\Gamma_R
    = 16.61 \sigma^* $ (the bistability threshold for $\sigma = 0.1 J$),
    $\Gamma_R = 11.52 \sigma^*$ (below the bistability threshold),
    and $27.12 \sigma^*$ (above the bistability threshold).
    (b) - The results obtained for
    $\Gamma_R = 16.61 \sigma^*$ and various detunings $\Delta_0$.
    The dotted and solid curves represent, respectively, the
    data calculated for $\Delta_0 = \Delta_0^{\mathrm{opt}} - \sigma^*$,
    and $\Delta_0^{\mathrm{opt}} + \sigma^*$.
    The open circles show the same data as in panel (a). }
    \label{Fig: Steady State}
\end{figure}

Several examples of the $S$-shaped input-output characteristics
calculated for the disorder degree $\sigma = 0.1 J$ are shown in
Fig.~\ref{Fig: Steady State} for an input field that is resonant
with the absorption maximum. We use the dimensionless intensities
$I_{\mathrm{in}} = |\Omega_i|^2/(\gamma_0 \sigma^*)$ and
$I_{\mathrm{out}} = |\Omega|^2/(\gamma_0 \sigma^*)$, which is
convenient because the HWHM of the absorption spectrum $\sigma^*$ is
an experimentally measurable quantity. Panel (a) shows how the
input-output characteristics change when $\Gamma_R$ is below, at, or
above its critical value. Panel (b) shows the input-output
characteristics when the field is tuned through the resonance.

\begin{figure}[ht]
\begin{center}
\includegraphics[width = 0.45\textwidth,scale=1.1]{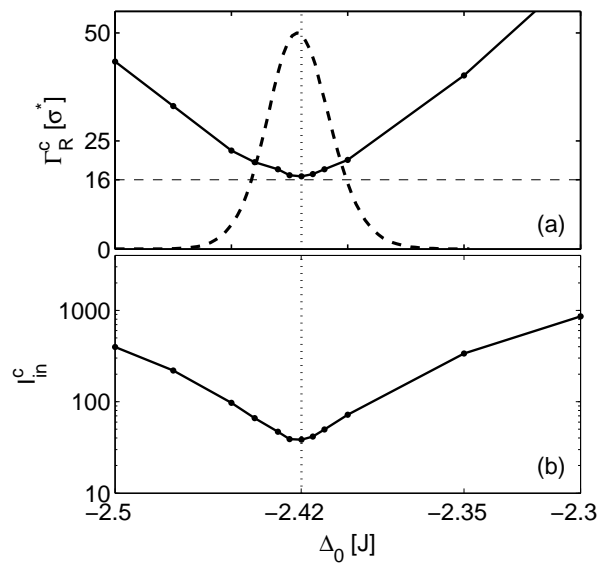}
\end{center}
\caption{(a) - Dependence of the critical superradiant constant
$\Gamma_R^c$ on the detuning $\Delta_0$ (solid line) calculated for
the disorder strength $\sigma = 0.1 J$. The dashed line shows the
absorption spectrum (absorption only due to $s$-states). The dotted
horizontal line indicates $\Gamma_R^c$ calculated for the optimal
detuning $\Delta^{\mathrm{opt}}_0 = -2.42  J$. (b) - Dependence of
the switching intensity $I_{\mathrm{in}}^c$ on the detuning
$\Delta_0$ calculated at the corresponding bistability threshold,
i.e., with $\Gamma_R^c$ given in the panel (a). } \label{Fig: Phase
diagram}
\end{figure}

At a given disorder strength $\sigma$, the minimal value of the
superradiant constant $\Gamma_R$ needed for optical bistability (the
\textit{critical} value $\Gamma_R^c$) depends on the detuning
$\Delta_0$. Figure~\ref{Fig: Phase diagram}(a) explicitly
demonstrates this effect: $\Gamma_R^c$ is almost constant within the
absorption band, whereas it clearly increases outside it. Panel (b)
shows the $\Delta_0$-dependence of the critical switching intensity
$I_{\mathrm{in}}^c$ of the incoming field at the bistability
threshold. This is the lowest intensity at which the film can
switch, when the field is tuned at $\Delta_0$, and when the
superradiance constant $\Gamma_R = \Gamma_R^c(\Delta_0)$. The data
presented here is obtained for the disorder strength $\sigma = 0.1
J$.

\begin{figure}[th]
\begin{center}
\includegraphics[width = 0.48\textwidth,scale=1.1]{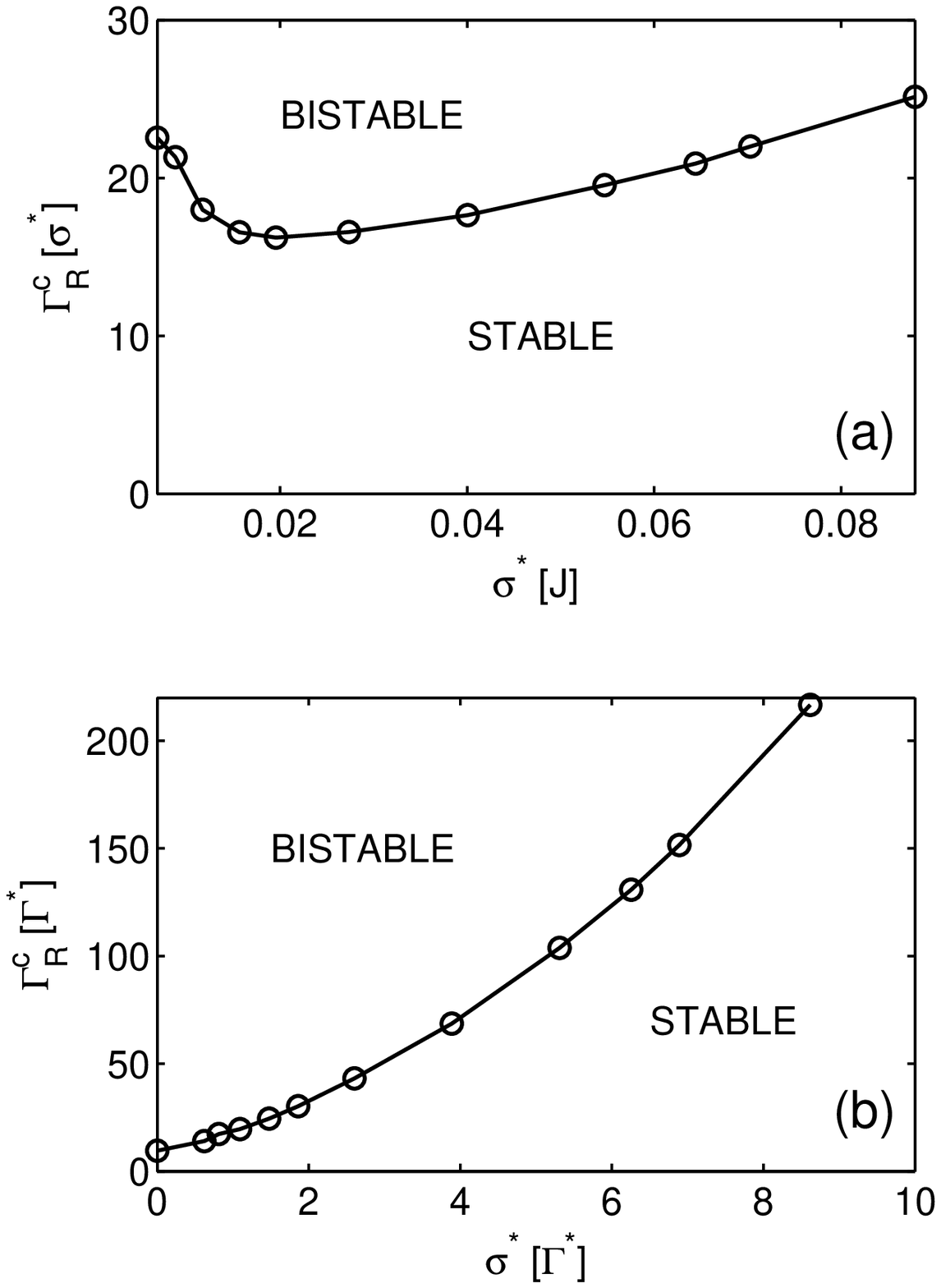}
\end{center}
\caption{(a) - Phase diagram of the bistable optical response of a
thin film in the $(\Gamma_R,\sigma^*)$-space obtained by solving
Eq.~(\ref{Eq: Steady state output field}) for
$\Delta=\Delta_0^{\mathrm{opt}}$. The open circles represent the
numerical data points, whereas the solid line is a guide to the eye.
Above (below) the solid line the film behaves in a bistable (stable)
fashion. The solid line itself represents the $\sigma^*$-dependence
of the critical superradiant constant $\Gamma_R^c$, calculated for
the optimal detuning $\Delta_0^{\mathrm{opt}}$, i.e., when the
incoming field is tuned to the absorption band maximum. This gives
the minimal $\Gamma_R^c$ for each $\sigma^*$. (b) - The same data
points as in the panel (a), only replotted as a function of
$\Gamma^*$, where $\Gamma^*$ is the mean value of the relaxation
constant $\Gamma$.} \label{Fig: Phase diagram sigma}
\end{figure}

As is seen from Fig.~\ref{Fig: Phase diagram}(a), there exists a
detuning $\Delta_0^{\mathrm{opt}}$, referred to as {\it the optimal
one}, at which $\Gamma_R^c$ takes its minimal value. The detuning is
optimal if the imaginary term in Eq.~(\ref{Eq: Steady state output
field}) vanishes: this term opposes a three-valued solution for the
output field. For a symmetric absorption band, the optimal detuning
corresponds to the incoming field being resonant with the absorption
maximum. In our case, owing to a small asymmetry of the absorption
band [see Fig.~\ref{fig: Example G}({\it b})],
$\Delta_0^{\mathrm{opt}} = - 2.42 J$  is shifted slightly to the
blue from the position of the absorption peak.

We calculated $\Gamma_R^c$ as a function of the HWHM $\sigma^*$ for
the optimal detuning. The result is shown in Fig.~\ref{Fig: Phase
diagram sigma}. The plot represents, in fact, the phase diagram of
the optical response: below the curve, the output-input
characteristic of the film is always single-valued (stable), while -
depending on the detuning - it can become three-valued (bistable)
above it. The nonmonotonic behavior of $\Gamma_R^c$ at small
magnitudes of $\sigma^*$, presented in the panel {\it a}, is simply
explained by the fact that the disorder-induced (inhomogeneous)
broadening becomes smaller than the homogeneous one $\sigma^* <
\Gamma^*$, where $\Gamma^*$ is defined as $\Gamma^* = \int d\mu
d\varepsilon \, \Gamma {\cal G}_s(\varepsilon,\mu)$. The ratio
$\Gamma_R/\Gamma^*$ is now the relevant parameter, governing the
occurrence of bistability. The panel ({\it b}) shows the
$\sigma^*$-dependence of $\Gamma_R^c$ replotted in units of
$\Gamma^*$, which is monotonic. When $\sigma \to 0$, the ratio
$\Gamma_R^c/\Gamma^* \to 9.64$. This value is deduced from
Eq.~(\ref{Eq: Steady state output field}). Indeed, in the limit of
$\sigma \to 0$ we can move the Lorenztian outside the integral and
use $\int d\Delta d\mu \, \mu^2 {\mathcal G}_s(\Delta,\mu) =
F_s/N_s$. The resulting equation is the same as for a thin film of
homogeneously broadened two-level systems, only with the
renormalized cooperative number ${\tilde \Gamma}_R/\Gamma^* =
(\Gamma_R/\Gamma^*)(F_s/N)$, where $F_s/N = 0.83$. Bearing in mind
that the critical value of the ratio ${\tilde \Gamma}_R/\Gamma^*$ is
equal to 8,~\cite{Zakharov88} we recover $\Gamma_R^c/\Gamma^* =
9.64$.

\subsection{Spectral distribution of the exciton population}
    \label{Sec: Spectral distribution}

More insight into what occurs at the switching threshold is
obtained by studying the population distribution
\begin{equation}
    \label{Eq: Population distribution}
   r_{11}(\Delta) = \int d\mu {\cal G}_s(\Delta,\mu)
   \rho_{11}(\Delta,\mu) \ ,
\end{equation}
with $\rho_{11}$ the steady-state solution of Eqs.~(\ref{Eq: Density
matrix truncated}). This distribution enables us to visualize the
relevant spectral range around the switching point.

\begin{figure}
\begin{center}
\includegraphics[width=0.48\textwidth]{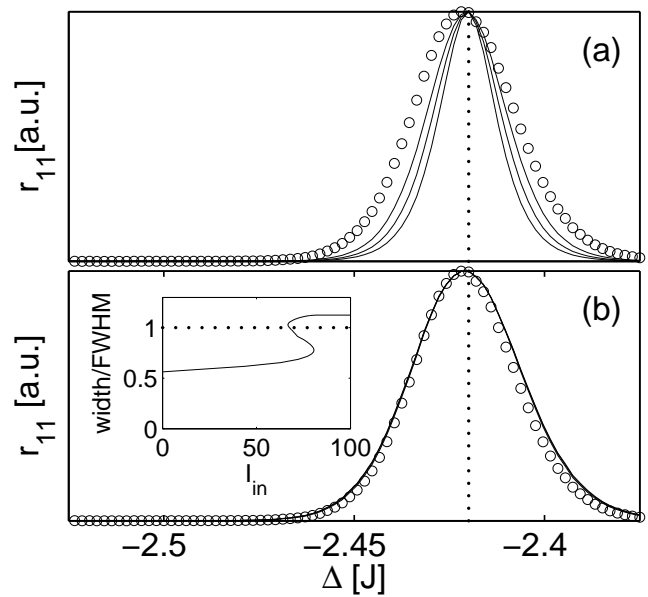}
\end{center}
 \caption{Population-distributions $r_{11}(\Delta)$
 (solid curves), calculated according to Eq.~(\ref{Eq: Population distribution})
 for $\sigma = 0.1J$ and $\Gamma_R = 27.12 \sigma^*$, with the optimal detuning
 $\Delta_0^{\mathrm{opt}} = -2.42 J$ indicated by the vertical dashed
 line. Open circles show the absorption spectrum $\mathcal{A}_s(\Delta)$.
 Panel (a) represents $r_{11}(\Delta)$ below the upper switching threshold.
 The plotted distributions were calculated for the input intensities
 $I_{\mathrm{in}} = |\Omega_i|^2/(\gamma_0\sigma^*) = 3.33, 64.34$, and
 $81.78$ (from bottom to top). Panel (b) shows  $r_{11}(\Delta)$ above the
 upper switching threshold. In the inset, the dependence of the full width
 at half maximum (FWHM) of $r_{11}(\Delta)$ on $I_{\mathrm{in}}$ is plotted
 in units of the FWHM of the absorption spectrum.} \label{Fig: Excited state
population}
\end{figure}

In Fig.~\ref{Fig: Excited state population}, we plotted
$r_{11}(\Delta)$ calculated for the optimal detuning
$\Delta_0^{\mathrm{opt}}$ and $\Gamma_R = 27.12 \sigma^*$ (above the
critical value $\Gamma_R^c$). Panels (a) and (b) show the results
obtained for the incoming field intensities $I_{\mathrm{in}} =
\Omega_{i}^2/(\gamma_0 \sigma^*)$  below and above the switching
threshold, respectively. Below the switching threshold, only a
relatively narrow spectral region around $\Delta^{\mathrm{opt}}_0$
acquires population. This is because, in spite of the intensities of
the incoming field $I_{\mathrm{in}} = 3.33, 64.34$, and $81.78$
being far above the saturation value, the intensity of the field
inside the film, $I_{\mathrm{out}} = \Omega_{i}^2/(\gamma_0
\sigma^*)$ $= 0.025$, $0.5$, and $1.5$, is below or on the order of
it. For these intensities, the one-exciton approximation, with only
one $s$-like excited state considered in each localization segment,
is reasonable.

Figure~\ref{Fig: Excited state population}(b) represents the
population distribution $r_{11}(\Delta)$ after switching, when the
field inside the film $I_{\mathrm{out}}$ exceeds the switching
threshold and becomes much larger than the saturation magnitude. In
this limit, we can replace $\rho_{11}(\Delta,\mu)$ in Eq.~(\ref{Eq:
Population distribution}) by 0.5 and get $r_{11}(\Delta) = 0.5 \int
d\mu {\cal G}_s(\Delta,\mu) = 0.5 {\cal D}_s(\Delta)$, where ${\cal
D}_s(\Delta)$ is the density of $s$-like states. The latter is
plotted in Fig.~\ref{Fig: Excited state population} (b) by the solid
line and appears to be wider than the absorption band. For such
field intensities, it is likely that the two-level model should be
corrected by including the one-to-two exciton transitions. This work
is now in progress.

\section{Time-domain analysis}
    \label{Sec: Time-domain analysis}

\subsection{Hysteresis loop}
    \label{Sec: Hysterisis loop}

\begin{figure}
\begin{center}
\includegraphics[width=0.48\textwidth]{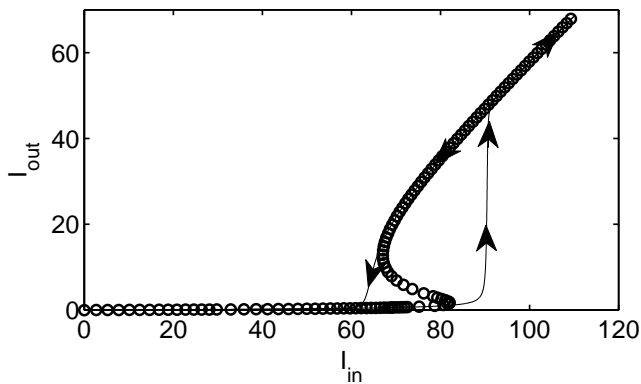}
\end{center}
\caption{An example of the stable optical hysteresis loop of the
transmitted intensity $I_{\mathrm{out}} =
|\Omega|^2/(\gamma_0\sigma^*)$ (the solid curve with arrows)
obtained by numerically  solving Eqs.~(\ref{Eq: Density matrix
truncated}) for a linear sweeping up-and-down of the input field
intensity $I_{\mathrm{in}} = |\Omega_i|^2/(\gamma_0\sigma^*)$. The
sweeping time is $3000/ \sigma^*$. The open circles represent the
steady-state solution, Eq.~(\ref{Eq: Steady state output field}).
The calculations were performed for the following set of parameters:
$\Gamma_2 = 500\gamma_0$, $\sigma = 0.1 J$, $\Gamma_R = 27.12
\sigma^*$, and $\Delta_0 = \Delta_0^{\mathrm{opt}} = -2.42 J$.}
\label{Fig: Hysteresis}
\end{figure}

It is well known that the S-shaped output-input dependence and, as a
consequence, the existence of two switching thresholds results in
optical hysteresis.~\cite{Lugiato84,Gibbs85} To investigate this, we
numerically integrated Eqs.~(\ref{Eq: Density matrix truncated})
while slowly sweeping up-and-down the input intensity
$I_{\mathrm{in}}$ above the bistability threshold ($\Gamma_R >
\Gamma_R^c$). The result for the transmitted intensity
$I_{\mathrm{out}}$ is shown in Fig.~\ref{Fig: Hysteresis} by the
solid curve with arrows. The parameters used in the calculations are
specified in the figure caption. The input field intensity was swept
from zero to 110 and back to zero. The open circles indicate the
steady-state solution obtained by solving Eq.~(\ref{Eq: Steady state
output field}) for the same set of parameters.

As can be seen from Fig.~\ref{Fig: Hysteresis}, the solid curve
almost perfectly follows the lower and upper branches of the
steady-state three-valued solution, nicely demonstrating the optical
hysteresis. The intermediate branch is not revealed, which is clear
evidence of its instability. Note also that switching from the lower
branch to the upper one occurs for an input field intensity larger
than the critical value. This indicates that when the input field
intensity is only slightly above the switching intensity, the
response of the film slows down. A much less pronounced but similar
effect can be observed at the lower switching threshold, where the
field switches from the upper branch to the lower one. This is
consistent with our study of the relaxation time presented below.

\subsection{Switching time}
    \label{Sec: Switching time}

Of great importance from a practical point of view, is the
relaxation time $\tau$ which is required for the output intensity to
approach its steady-state value after the input intensity has
changed. If this time is much shorter than the characteristic time
of changing the input intensity, then the output signal will
adiabatically follow it, remaining all the time close to the
steady-state level. Only in the limit of short $\tau$, an abrupt
switching from low to high transmittivity can be realized. This
especially concerns the region in the vicinity of the switching
thresholds (see Fig.~\ref{Fig: Hysteresis}). In other words, the
relaxation time $\tau$ limits the usage of the optical bistable
element as an instantaneous switcher.

\begin{figure}
\begin{center}
\includegraphics[width=0.48\textwidth]{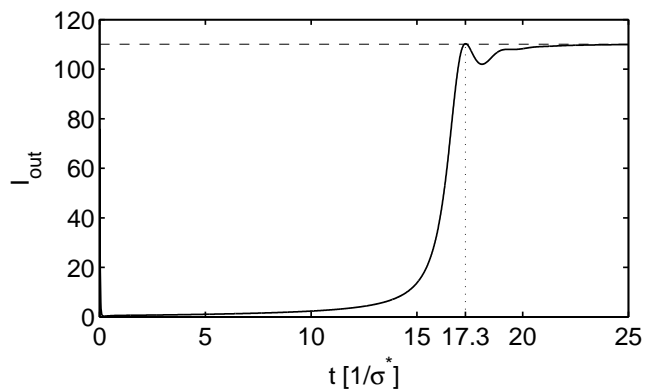}
\end{center}
\caption{Kinetics of the transmitted field
    intensity $I_{\mathrm{out}} = |\Omega|^2/(\gamma_0\sigma^*)$
    approaching its stationary value (dashed line) after the incident field with
    intensity $I_{\mathrm{in}}= |\Omega_i|^2/(\gamma_0\sigma^*) = 150$
    is turned on abruptly at $t = 0$. The value $I_{\mathrm{in}} = 150$ exceeds the
    upper switching threshold $I_{\mathrm{in}}^c = 82.16$. The other
    parameters were chosen as in Fig.~\ref{Fig: Hysteresis}.}
\label{Fig: Example Switching}
\end{figure}

Motivated by the above observations, we performed a study of  the
relaxation time $\tau$. Figure~\ref{Fig: Example Switching} shows an
example of how the transmitted field intensity approaches its
stationary value when an input field intensity with a value of
$I_{\mathrm{in}} = 150$ is instantaneously switched on at $t=0$.
This field is above the upper switching threshold $I_{\mathrm{in}}^c
= 82.16$. Calculations were carried out for the set of parameters of
Fig.~\ref{Fig: Hysteresis}. As is observed, for the set of
parameters used, the output intensity stays low during a waiting
time of about $20/\sigma^*$, before it rapidly (on a time scale much
shorter than $20/\sigma^*$) increases to its steady state value.
This behavior allows one to define $\tau$ as the time which the
output intensity takes to reach its first peak ($17.3/\sigma^*$ in
the current example).

\begin{figure}
\begin{center}
\includegraphics[width=0.48\textwidth]{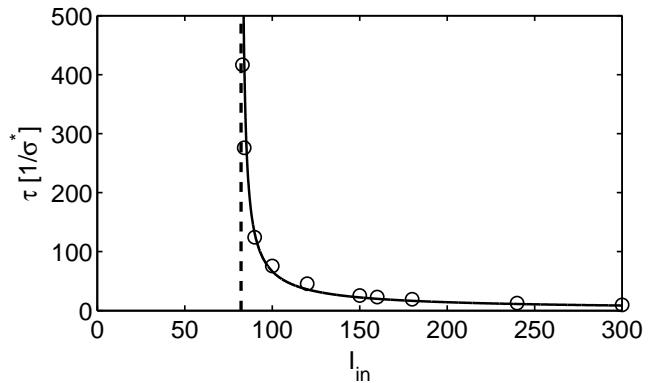}
\end{center}
\caption{Relaxation time $\tau$ as a function of the excess input
    intensity $I_{\mathrm{in}} - I_{\mathrm{in}}^c$ at
    the upper switching threshold (indicated by the vertical dotted
    line). $\tau$ was calculated by turning on abruptly the incoming
    field at $t=0$, and waiting until the transmitted field intensity
    $I_{\mathrm{out}}$ approaches its steady-state value (for more
    details, see the text).
    The open circles show the numerical results, while the solid
    line represents a best power-law fit given by Eq.~\ref{Eq: Relaxation time}.
    The calculations were performed for the set of parameters
    of Fig.~\ref{Fig: Hysteresis}.}
\label{Fig: Switching time versus I}
\end{figure}

Using the above definition, we calculated the relaxation time $\tau$
as a function of the excess  input intensity $I_{\mathrm{in}} -
I_{\mathrm{in}}^c$ at the upper switching threshold. The results are
plotted in Fig.~\ref{Fig: Switching time versus I}. As is seen,
$\tau$ drastically increases when $I_{\mathrm{in}}$ gets closer to
$I_{\mathrm{in}}^c$. The numerical data (open circles) is well
fitted by the formula
\begin{equation}\label{Eq: Relaxation time}
     \tau = 870
 \left(I_{\mathrm{in}}-I_{\mathrm{in}}^c\right)^{-0.83} \ ,
\end{equation}
shown by the solid curve.

\section{Discussion of driving parameters}
    \label{Sec: Driving Parameters}

To get insight into the possibility to  realize optical bistable
behavior for a film of J-aggregates, we  consider the typical
parameters for this type of systems. First, we estimate the
superradiant constant $\Gamma_R = (3/8\pi) \gamma_0 n_0 \lambda^2
L$, considering the low-temperature experimental data of
J-aggregates of polymethine dyes. For these species, typically,
$\gamma_0 \approx (1/3)$ ns$^{-1}$ and $\lambda \approx 600$
nm.~\cite{deBoer90,Fidder90,Minoshima94,Moll95,Kamalov96,Scheblykin00}
With this in mind and choosing $L = \lambda/2\pi$ \ (or $kL = 1$),we
obtain the following estimate: $\Gamma_R \approx  10^{-18} n_0$
cm$^3$ ps$^{-1}$. This value for $L$ is easily accessible with the
spin-coating method~\cite{Misawa93} and guarantees the applicability
of the mean-field approach for the description of the thin film
optical response.~\cite{Jarque01} The typical width of the J-band of
polymethine dyes is on the order of several tens of cm$^{-1}$ or
approximately 1 ps$^{-1}$ (in frequency
units).~\cite{deBoer90,Fidder90,Minoshima94,Moll95,Kamalov96,Scheblykin00}
Thus, for the set of parameters we chose, the number density of
molecules $n_0$ must be on the order of $10^{19}$ cm$^{-3}$ to get
the ratio $\Gamma_R/\sigma^*$ required for bistability to occur.
This concentration is usually achieved in spin-coated films.

Another option to adjust the parameters favoring bistability is to
consider $J$-aggregates composed of monomers with higher radiative
constant $\gamma_0$ and a larger emission wavelength $\lambda$. From
this point of view, J-aggregates of squarylium dyes may be suitable
candidates.~\cite{Furuki01,Tatsuura01,Pu02} This type of aggregates,
spin-coated on a substrate, shows a sharp absorption peak at
$\lambda \approx 800$ nm with HWHM = 20 nm at room temperature and a
fast ($\sim 100$ fs) optical response~\cite{Furuki01,Pu02} combined
with a giant cubic succeptibility,~\cite{Tatsuura01} both attributed
to the excitonic nature of the optical excitations. The monomer
decay time has been reported to be $\sim 100$ ps,~\cite{Furuki01}),
although no information about the quantum yield has been presented.
If we assume that this time is of radiative nature, the superradiant
constant $\Gamma_R$ can be adjusted to values above the bistability
threshold even for smaller concentration of monomers in the film. On
the other hand, for larger $\gamma_0$ also the intensity required
for switching increases, which is not desired because of the limited
photostability of most J-aggregates.



\section{Summary and concluding remarks}
    \label{Sec: Summary}

We theoretically studied the optical response of an ultrathin film
of oriented J-aggregates, with the goal to examine the possibility
of bistable behavior of the system. The standard Frenkel exciton
model was used for a single aggregate: an open linear chain of
monomers coupled by delocalizing dipole-dipole excitation transfer
interactions, in combination with uncorrelated on-site disorder,
which tends to localize the exciton states. We considered a single
aggregate as a meso-ensemble of two-level systems, each one composed
of an $s$-like localized one-exciton state and its own ground state.
The one-to-two exciton transitions have been neglected.

As a tool to describe the transmission properties of the film, we
employed the optical Maxwell-Bloch equations adapted for a thin
film. The electric polarization of the film was calculated by making
use of a joint probability distribution of exciton energies and
transition dipole moments, properly taking into account the
correlation properties of these two stochastic variables. The joint
distribution function was calculated by numerically diagonalizing
the Frenkel Hamiltonian and averaging over many disorder
realizations.

We derived a novel steady-state equation for the transmitted signal
in terms of the joint distribution function, and demonstrated that
three-valued solutions to this equation exist in a certain domain of
the parameter space $(\Gamma_R,\sigma^*)$,where $\Gamma_R$ is the
superradiant constant and $\sigma^*$ is the
half-width-at-half-maximum of the absorption band. Our approach
allowed us to generalize previous results~\cite{Malyshev00,Jarque01}
to correctly account for the stochastic nature of the exciton energy
and transition dipole moment. Using the new steady-state equation,
we have found that the critical value of the so-called "cooperative
number" $\Gamma_R/\sigma^*$,~\cite{Lugiato84} which governs the
occurrence of bistability of the film, is higher than obtained
before.~\cite{Malyshev00} Moreover, in contrast to
Refs.~\onlinecite{Malyshev00} and~\onlinecite{Jarque01}, we have
analyzed the switching time, which show a dramatic increase for
input intensities close to the switching point. We also found that
the "cooperative number" $\Gamma_R/\sigma^*$ increases with
$\sigma^*$, but only slightly, varying between 12 and approximately
25 within a wide range of $\sigma^*$. Estimating the parameters of
our model for aggregate of polymethin dyes shows that these are a
promising candidate to measure the effect.

Finally, we note that also the microcavity arrangement of molecular
aggregates~\cite{Lidzey98,Litinskaya04,Beltyugov04,Agranovich05,Zoubi05}
is of interest for applications. During the last decade, organic
microcavities have received a great deal of attention because of the
strong coupling of the excitons to cavity photons, leading to giant
polariton splitting in these devices.~\cite{Lidzey03} The recent
observation of optical bistability in planar {\it inorganic}
microcavities~\cite{Baas04} in the strong coupling regime suggests
that {\it organic} microvavities can exhibit a similar behavior.


\acknowledgments This work is part of the research program of the
Stichting voor Fundamenteel Onderzoek der Materie (FOM), which is
financially supported by the Nederlandse Organisatie voor
Wetenschappelijk Onderzoek (NWO). Support was also received from
NanoNed, a national nanotechnology programme coordinated by the
Dutch Ministry of Economic Affairs.

\appendix

\section{Estimates of quantum interference effects}

Our approach to the optical dynamics of a single aggregate was based
on the representation of the aggregate as a meso-ensemble of
two-level systems with their own ground states. The model has its
origin in the fact that the optically dominant exciton states are
localized on different segments and overlap weakly. In reality,
however, the ground state of an aggregate (all the monomers are in
their ground states) is common for all (multi-~) exciton states.
This results in cross-interference of field-induced as well as
spontaneous transitions. Below, we provide estimates of these
additional terms and show that in the limit of dominant
inhomogeneous broadening of the J-band, the cross-interference
effects can be neglected. In our estimates, we will only consider
ground state-to-one-exciton transitions.

We start with the equation for the density operator $\rho$
\begin{equation}
    \label{rho}
    \dot{\rho}
    = -\frac{i}{\hbar}\left[ H_0 - {\hat d}\mathcal{E}, \rho \right]
    -R^{\mathrm{bath}} \rho - R^{\mathrm{rad}} \rho \ ,
\end{equation}
where $H_0$ is the exciton Hamiltonian specified in Eq.~(\ref{H})
and the term $-{\hat d}\mathcal{E}$ describes the interaction of the
aggregate with the field $\mathcal{E}$ inside the film.
$R^{\mathrm{bath}}$ represents the dephasing operator, acting as
follows:
\begin{eqnarray}
    \label{Rbath}
    \langle \nu| R^{\mathrm{bath}}\rho \, |\nu^{\prime}\rangle
    & = & (1 - \delta_{\nu\nu^{\prime}}) (\gamma_{\nu 0}
    + \gamma_{\nu^{\prime}0}) \rho_{\nu\nu^{\prime}}\ ,
\nonumber\\
\nonumber\\
    \langle \nu| R^{\mathrm{bath}}\rho \, |0 \rangle
    & = & \gamma_{\nu 0}\rho_{\nu 0} \ .
\end{eqnarray}
\\
Here, $\gamma_{\nu 0}$ is the (pure) dephasing rate of the exciton
transition $|\nu \rangle \to |0 \rangle$, excluding radiative
decay. These constants will be considered on a phenomenological
basis.

The operator $R^{\mathrm{rad}}$ governs the exciton radiative
relaxation. It is given by (see, e.g., Ref.~\onlinecite{Blum96})
\begin{eqnarray}
    \label{Rrad}
    R^{\mathrm{rad}} \rho
    & = & \frac {1}{2}\> \sum_{\nu\nu^{\prime}} \gamma_{\nu\nu^{\prime}}
    \Big [\, |\nu \rangle \langle \nu^{\prime}|\, \rho
\nonumber\\
    & + & \rho \, |\nu \rangle \langle \nu^{\prime}|
    - 2\, |0 \rangle \langle \nu|\, \rho \,|\nu^{\prime} \rangle
    \langle 0|\> \Big ]\ ,
\end{eqnarray}
where $\gamma_{\nu\nu} = \gamma_0 \big( \sum_n \varphi_{\nu n}
\big)^2$ is the radiative decay rate of the population of the
$\nu$th state. Furthermore, $\gamma_{\nu\nu^{\prime}} =
\gamma_{\nu^{\prime}\nu}$ \, ($\nu \ne \nu^{\prime}$) describes the
quantum interference in the radiative relaxation of the $\nu$th and
$\nu^{\prime}$th states, resulting from the cross-coupling of
different decay channels. It reflects the fact that a state $\nu$,
when decaying, drives another state $\nu^{\prime}$ and {\it vice
versa}. If the transition dipoles of all the states are parallel,
$\gamma_{\nu\nu^{\prime}} = (\gamma_{\nu\nu}
\gamma_{\nu^{\prime}\nu^{\prime}})^{1/2}$.

Using Eqs.~(\ref{Rbath}) and~(\ref{Rrad}) in Eq.~(\ref{rho}), we
arrive at the following set of equations for the density matrix
elements:
\begin{widetext}
\begin{subequations}
\label{Density_matrix_equations}
\begin{equation}
    \label{rho-nunu}
    \dot{\rho}_{\nu\nu} = - \gamma_{\nu\nu} \rho_{\nu\nu}
    - \frac{1}{2}\sum_{\nu^{\prime}\ne\nu} \gamma_{\nu\nu^{\prime}}
    \left(\rho_{\nu\nu^{\prime}} + \rho_{\nu^{\prime}\nu}\right)
    +i d_{\nu}\mathcal{E}\left(  \rho_{\nu 0}^*
    - \rho_{\nu 0} \right) \ ,
\end{equation}
\begin{equation}
    \label{rho-nunu'}
    \dot{\rho}_{\nu\nu^{\prime}} = - (i\varepsilon_{\nu\nu^{\prime}}
    + \Gamma_{\nu\nu^{\prime}})\rho_{\nu\nu^{\prime}}
    - \frac{1}{2}\sum_{\nu^{\prime\prime}\ne\nu}
\gamma_{\nu\nu^{\prime\prime}}
    \rho_{\nu^{\prime\prime}\nu^{\prime}}
    - \frac{1}{2}\sum_{\nu^{\prime\prime}\ne\nu^{\prime}}
    \gamma_{\nu^{\prime\prime}\nu^{\prime}}
    \rho_{\nu\nu^{\prime\prime}}
    +i \left( d_{\nu}\mathcal{E} \rho_{\nu^{\prime} 0}^*
    - \rho_{\nu 0} d_{\nu^{\prime}}\mathcal{E} \right) \ ,
    \quad\quad \nu \ne \nu^{\prime} \ ,
\end{equation}
\begin{equation}
    \label{rho-nu0}
    \dot{\rho}_{\nu 0} = - (i\varepsilon_{\nu} + \Gamma_{\nu 0})\rho_{\nu 0}
    - \frac{1}{2}\sum_{\nu^{\prime}\ne\nu} \gamma_{\nu\nu^{\prime}}
    \rho_{\nu^{\prime}0}
    -  i \sum_{\nu^{\prime} \ne \nu} \rho_{\nu \nu^{\prime}}
    d_{\nu^{\prime}}\mathcal{E}
    - i d_{\nu}\mathcal{E}\left( \rho_{\nu \nu}
    - \rho_{00} \right) \ ,
\end{equation}
\begin{equation}
    \label{rho-00}
    \rho_{00} + \sum_{\nu} \rho_{\nu \nu}  = 1 \ .
\end{equation}
\end{subequations}
\end{widetext}
Here we introduced the notations: $\varepsilon_{\nu\nu^{\prime}} =
\varepsilon_{\nu} - \varepsilon_{\nu^{\prime}}$,
$\Gamma_{\nu\nu^{\prime}} = \frac{1}{2}( \gamma_{\nu\nu} +
\gamma_{\nu^{\prime}\nu^\prime}) + \gamma_{\nu 0} +
\gamma_{\nu^{\prime} 0}$, and $\Gamma_{\nu 0} = \frac{1}{2}
\gamma_{\nu\nu} + \gamma_{\nu 0}$.

Equations~(\ref{Density_matrix_equations}) differ from those used in
the two-level model, Eq.~(\ref{Eq: Density matrix}), by several
terms. Because all the exciton states have the same ground state,
which is reflected in the normalization condition~(\ref{rho-00}),
the low-frequency coherences are now involved in the aggregate
optical dynamics. They are coupled to the populations
[Eq.~(\ref{rho-nunu})] as well as to the high-frequency (optical)
coherences $\rho_{\nu 0}$ [Eqs.~(\ref{rho-nunu'})
and~(\ref{rho-nu0})] via both the cross-coupling of the transitions
and the field. In addition, the cross-coupling also couples the
optical coherences $\rho_{\nu 0}$ [Eq.~(\ref{rho-nu0})].

In quantum optics of atomic gases, the cross-coupling of transitions
has been found to be the origin of many fascinating effects, such as
narrow resonances, transparency and gain without population
inversion (see for an overview
Refs.~\onlinecite{Kocharovskaya92,Arimondo96,Harris97,Fleischhauer05}),
as well as bistability at a low
threshold.~\cite{Walls80,Anton02,Joshi04} All these effects,
however, require specific conditions: equivalent magnitudes of all
the $\gamma_{\nu\nu}$ and the absence of dephasing and inhomogeneous
broadening. Any deviation from these requirements washes out those
effects. In particular, this  happens for J-aggregates; below we
argue why all the cross-terms in
Eqs.~(\ref{Density_matrix_equations}) can be neglected for these
systems.

The contribution of the cross-terms to a given state $\nu$ always
comes in the form of a summation over all other states
$\nu^{\prime}$. The optical dynamics of the system is determined by
only several dominant states. If $N^*$ is the typical localization
length, there will be $N/N^*$ of such states. They are spread over
the width of the absorption band, given by $2\sigma^*$. Therefore we
can estimate the sum under consideration by
$(\gamma^*/2\sigma^*)(N/N^*) \sim \gamma_0 N/2\sigma^*$, where
$\gamma^* = \gamma_0 N^*$ is the typical radiative rate of optically
dominant states. The materials we consider typically have $\gamma_0
\sim 10^{8}$ s$^{-1} \sim 10^{-2}$ cm$^{-1}$ and $2\sigma^*$ on the
order of several tens of cm$^{-1}$. Then, for an aggregate of length
$N = 500$ the ratio $\gamma_0 N/2\sigma^* \sim 0.1$. On this basis,
we neglect all the cross-coupling terms in
Eqs.~(\ref{Density_matrix_equations}) and replace the normalization
condition~(\ref{rho-00}) for the whole aggregate by the one for a
single segment, $\rho_{00} + \rho_{\nu\nu} = 1$.


\end{document}